\documentclass[aps,reprint,superscriptaddress,prl,twocolumn,showpacs]{revtex4-1}

\usepackage{epsfig}
\usepackage{amsmath}
\usepackage{amssymb}
\usepackage{color}
\usepackage{bm}

\newcommand{\ket}[1]{\left|#1\right\rangle}
\newcommand{\bra}[1]{\left\langle#1\right|}
\usepackage{verbatim}
\definecolor{purple}{RGB}{160,32,240}
\newcommand{\TK}[1]{#1}

\definecolor{seegroen}{rgb}{0,0.5,0.5}

\newcommand{\strike}[1]{\textcolor{red}{*}}

\begin{document}
%=================================================================

\title{Implementing Quantum Walks Using Orbital Angular Momentum of Classical Light}
%-------------------------------------

\author{Sandeep K \surname{Goyal}}
\email{goyal@ukzn.ac.za}
\affiliation{School of Chemistry and Physics, University of KwaZulu-Natal, Private Bag X54001, Durban 4000, South Africa}  
\affiliation{Center for Quantum Sciences, The Institute of Mathematical Sciences, CIT Campus, Chennai 600 113, India}
%------------------------------------

\author{Filippus S \surname{Roux}}
\affiliation{CSIR National Laser Centre, PO Box 395, Pretoria 0001, South Africa}

\author{Andrew \surname{Forbes}}
\affiliation{CSIR National Laser Centre, PO Box 395, Pretoria 0001, South Africa}
\affiliation{School of Chemistry and Physics, University of KwaZulu-Natal, Private Bag X54001, Durban 4000, South Africa}

\author{Thomas \surname{Konrad}}
\affiliation{School of Chemistry and Physics, University of KwaZulu-Natal, Private Bag X54001, Durban 4000, South Africa}  
\affiliation{National Institute for Theoretical Physics (NITheP), University of KwaZulu-Natal, Private Bag X54001, Durban 4000, South Africa}

%===========================================
\begin{abstract}
We present an implementation scheme for a quantum walk in the orbital angular momentum space of a laser beam. The scheme makes use of a ring interferometer, containing a quarter-wave plate and a $q$ plate. This setup enables one to perform an arbitrary number of quantum walk steps. In addition, the classical nature of the implementation scheme makes it possible to observe the quantum walk evolution in real time. We use nonquantum entanglement of the laser beam's polarization with its  orbital angular momentum to implement the quantum walk.
\end{abstract}

%03.67.-a Quantum information
%42.50.-p Quantum optics
%42.50.Ex Optical implementations of quantum information processing and transfer
%42.50.Tx Optical angular momentum and its quantum aspects
%42.25.Ja Polarization
%42.25.Lc Birefringence
%05.40.Fb Random walks and Levy flights
%03.65.Ta Foundations of quantum mechanics; measurement theory

%===========================================
\maketitle
%---------------------------------

Propagation by a succession of random steps is known as {a classical random walk}, where  the position of the walker is described by a Gaussian probability distribution after several random steps. Both discrete time quantum walks (DTQW) and classical random walks involve a random choice (coin toss) and a conditional drift (propagation),  but in a quantum walk the superposition of states allows interference of different paths, leading to strikingly different output distributions. For example, the spread of the probability distribution for the quantum walker increases quadratically 
%as fast as that of 
compared to
its classical counterpart \cite{Aharonov1993,Nayak2000,Ambainis2001,Kempe2003,Chandrashekar2008}. 

This speed up gained in quantum walks promises advantages when applied in quantum computation for certain 
%classes of 
quantum algorithms \cite{Ambainis2003}, 
%for example, 
such as quantum search algorithms \cite{Shenvi2003,Childs2004}. Quantum walks have also been used to analyze energy transport in biological systems \cite{Mohseni2008}.

Several experimental implementations of quantum walks have been reported over the years, using either atomic \cite{Karski2009,Zahringer2010,Schmitz2009,Xue2009} or photonic systems \cite{Souto2008,Bouwmeester1999,Broome2010,Zhang2010,Peruzzo2010}. Remarkably, none of these experiments could implement more than a few steps of the walker. The reason is that atomic systems require formidable control and isolation from their environment, which is difficult to achieve over many steps. Photonic quantum walks on the other hand, tend not to be scalable because the number of optical components (beam splitters and wave plates), which in practice induce losses and decoherence, grows too fast with the number of iterations in the quantum walk. Moreover, photonic quantum walks in general suffer from low efficiencies of single photon sources and detectors. However, it is possible to reduce the number of optical components drastically by using a loop in the experimental setup. Such a setup was used \cite{Schreiber2010,Schreiber2011} to demonstrate quantum walk over 28 steps in the time domain with attenuated laser pulses to simulate single photons.

%{ It is however, possible to reduce the number of optical components {\bf drastically} by {\bf using a loop in the experimental setup} \cite{Schreiber2010,Schreiber2011}. With this setup quantum walk over 28 steps in the time domain {\bf was demonstrated with} an attenuated laser pulse simulating a single photon.}

In this Letter we suggest a scalable photonic implementation scheme for DTQW. In our scheme { orbital angular momentum} (OAM) modes \cite{Allen1992,Leach2002} serve as the lattice sites and polarization is used to simulate the coin toss. A quarter-wave plate (QWP) \cite{Hecht2001} implements the coin flip operation and a $q$ plate \cite{Marrucci2006} generates conditional shifts in OAM space. Together they preform a single step of the quantum walk \TK{within a single beam without the need to split the beam and interfere light from different spatial paths} as, e.g.,\ in Refs. \cite{Schreiber2010,Schreiber2011}, which may add instability to the system. Since both optical devices work identically on the classical level (for laser beams) and on the quantum level (for single photons), one can apply the same scheme to realize a quantum walk with classical light, as proposed below. The classical implementation of the quantum walk is superior to the quantum realization as (i) it is easier to implement  and (ii) noninvasive measurements of classical systems enable a monitoring of the quantum walk in real time, { with a frame rate that depends on the detection device. For single photons,  the record of a single frame in the evolution requires many repetitions of the experiment to gather the probability distribution of the walker}.    Moreover, this implementation sheds light on the new concept of { nonquantum entanglement} \cite{Simon2010,Karimi2010, Gabriel2011, Holleczek2011, Qian2011, Marrucci2012}. 

A different realization scheme for a quantum walk with classical light using an electro-optic modulator has been proposed by Knight {\em et al.} \cite{Knight2003}. 
%An experimental quantum walk setup, using beam splitter arrays to distribute laser beams, is described in \cite{Jeong2004}, but this scheme is not scalable.
% \TK{Omitted one reference.} 
%{ A scalable quantum walk without coin (similar to a Galton board {\bf [Ref?]} ) using time bins of bright laser pulses in an optical fibre network was experimentally realized for 70 steps \cite{Schreiber3}.}
An experimental setup comprising an optical fibre network with time bined bright laser pulses has been used to implement a scalable coinless quantum walk, similar to a Galton board \cite{Schreiber3}. 

%In Section \ref{QW} we review the concept  of DTQW  which is employed in Section \ref{single-photon} to show that a quantum walk in OAM space can be realized with a single photon. The same optical devices used for this purpose --- a quarter-wave plate and a $q$ plate --- also give rise to a quantum walk with laser beams as demonstrated in Section \ref{classical}, followed by  a detailed implementation scheme in Section \ref{imple}. Although the laser light  shows characteristic features of a {\em coined} quantum walk, which resemble entanglement, we prove in  Section \ref{coherent}  that in fact there is no quantum entanglement present at any stage of the quantum walk with laser beams.

%\section{Review of quantum walk}\label{QW}

\begin{figure}
\includegraphics[width=8.5cm]{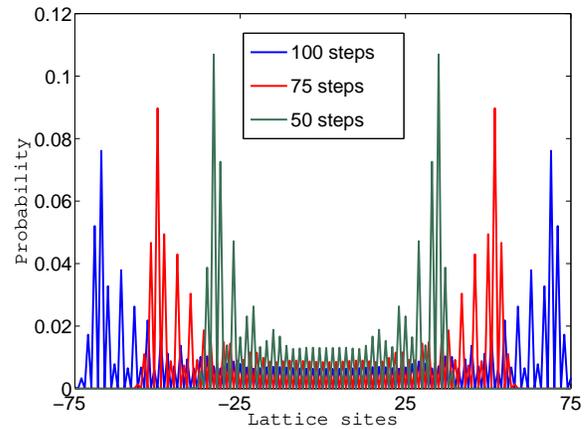}
\caption{(Color online) A typical evolution of the probability in a quantum walk on a one-dimensional lattice. The initial coin state of the particle is $(\ket{\uparrow} + i\ket{\downarrow})/\sqrt{2}$}.\label{fig-quantum-walk}
\end{figure}

We start with the description of DTQW in one spatial dimension. Let us represent the basis vectors of the coin space by $\{\ket{\uparrow},\ket{\downarrow}\}$ and the  basis of  the position space as $\{\ket{j}\}_{j=-\infty}^{\infty}$.% There are many ways to generalize the coin flip operation \cite{Chandrashekar2008}. 
One  way to define a quantum process that resembles the coin toss operation is to produce a weighted superpositions of heads ($\ket{\uparrow}$) and tails ($\ket{\downarrow}$):
\begin{align}
\ket{\uparrow} &\to \frac{1}{\sqrt{2}}\left( \ket{\uparrow} + \ket{\downarrow}\right),\\
\ket{\downarrow} &\to \frac{1}{\sqrt{2}}\left( \ket{\uparrow} - \ket{\downarrow}\right),
\end{align}
which corresponds to the Hadamard operator
\begin{align}
H &=  \frac{1}{\sqrt{2}}\left( \ket{\uparrow} + \ket{\downarrow}\right)\bra{\uparrow}+\frac{1}{\sqrt{2}}\left( \ket{\uparrow} - \ket{\downarrow}\right)\bra{\downarrow}.
\label{Hadamard}
\end{align}
As with classical random walk, DTQW require a conditional shift operation: shift of the position to the left (right) for tails (heads), which is represented by the shift operator
\begin{align}
S &= \sum_j \left(\ket{j+1}\bra{j} \otimes \ket{\uparrow}\bra{\uparrow} + \ket{j-1}\bra{j}\otimes \ket{\downarrow}\bra{\downarrow}\right).
\label{shift}
\end{align}
The quantum walk is realized by the iteration of the combined coin flip $H$ and shift $S$
\begin{align}
\tilde{H} S := (\mathbb{I}\otimes H)S 
=& \sum_j \left[ \ket{j+1}\bra{j} \otimes \left( {\ket{\uparrow}+\ket{\downarrow}\over\sqrt{2}} \right)\bra{\uparrow} \right.\nonumber \\ & \left.+ \ket{j-1}\bra{j}\otimes\left( {\ket{\uparrow}-\ket{\downarrow}\over\sqrt{2}} \right)\bra{\downarrow}\right] .
\label{hsexp}
\end{align} 
The combined operation causes a flip in the coin state along with a state dependent propagation on the lattice. For example, if the particle is initially at the lattice site $\ket{0}$ with spin state $(\ket{\uparrow} + i\ket{\downarrow})/\sqrt{2}$, the action of the operator $\tilde{H}S$ will map it to the state 
\begin{align}
\ket{\psi} = \frac{1}{2}\left[\ket{1}\otimes (\ket{\uparrow} + \ket{\downarrow}) + i\ket{-1}\otimes (\ket{\uparrow} - \ket{\downarrow})\right]. 
\label{entanglement1}
\end{align} 
In Fig.~\ref{fig-quantum-walk} we plot the probability distribution of the particle on the lattice for different numbers of steps.

%\section{Quantum walk for single photons}
%\label{single-photon}

%We now show that a 
A quantum walk can be realized using a photon that is transferred between different OAM modes, conditioned on its polarization by means of a $q$ plate. Consider a photon in the OAM eigenstate $\ket{\ell}$ and a superposition of left-handed and right-handed circular polarization $(\ket{L} + \ket{R})/\sqrt{2}$. The action of a $q$ plate on this state is given by
\begin{align}
\frac{1}{\sqrt{2}}(|L,\ell\rangle + |R,\ell\rangle) \rightarrow \frac{1}{\sqrt{2}}(|R,\ell-2q\rangle + |L,\ell+2q\rangle) ,
\end{align}
where $q$ is a fixed dimensionless parameter (a half integer) associated with the $q$ plate (cf.\ explanation of its working principle below). Note that the change in the OAM is twice the value of $q$. The $q$ plate acts qualitatively in a similar way as the conditional shift operator $S$ as in Eq.~(\ref{shift}):
\begin{align}
Q = \sum_{\ell=-\infty}^{\infty} & \left(\ket{\ell+2q}\bra{\ell} \otimes \ket{L}\bra{R} \right. \nonumber \\
& \left. + \ket{\ell-2q}\bra{\ell} \otimes \ket{R}\bra{L}\right)\,.\label{eqn-6}
\end{align}
We can realize the quantum walk evolution operator $\tilde{H}S$ by concatenating a $q$ plate and a (QWP). A quarter-wave plate with its optic axis rotated by an angle of $\theta = \pi/4$ is represented by the operator %, performs the following action on the polarization state \cite{Hecht2001} 
\begin{align}
W\left(\frac{\pi}{4}\right)& = \frac{1}{\sqrt{2}}(\ket{R}-\ket{L})\bra{R} + \frac{1}{\sqrt{2}}(\ket{R}+\ket{L})\bra{L}.
\end{align}
Thus, the product of the rotated quarter-wave plate and the $q$ plate yields
\begin{align}
W\left(\frac{\pi}{4}\right)Q = &\sum_{\ell=-\infty}^{\infty}\left[\ket{\ell+2q}\bra{\ell} \otimes \left(\frac{\ket{R} +\ket{L}}{\sqrt{2}}\right)\bra{R}\right.\nonumber\\ 
&\left. + \ket{\ell-2q}\bra{\ell} \otimes \left(\frac{\ket{R} - \ket{L}}{\sqrt{2}}\right)\bra{L}\right] ,
\end{align}
which is identical to the quantum walk evolution operator $\tilde{H}S$, given in Eq.~(\ref{hsexp}), and leads to the state in Eq.~(\ref{entanglement1}), provided that we set $q=1/2$ and identify the circular polarization states $\ket{L}$ and $\ket{R}$ with the basis for the coin space $\ket{\uparrow}$ and $\ket{\downarrow}$, respectively. Thus, we can realize a quantum walk in the space of OAM.

%\section{Quantum walk for laser beams}
%\label{classical}

However, it is not necessary to use single photons to obtain a quantum walk. In fact,  a simple laser pulse undergoes analogous dynamics when subjected to a $q$ plate followed by a quarter wave plate. This can be easily understood in the language of classical optics using Jones matrices \cite{Jones1941a, Hurwitz1941, Jones1941b, Jones1942}, which yield a remarkably concise representation of the quantum walk.

%We now discuss the light modes of a laser beam, the action of the quarter-wave plate and the working principle of the $q$ plate in the classical picture, in terms of the Jones matrix formalism \cite{Jones1941a, Hurwitz1941, Jones1941b, Jones1942}. Our aim is to show that a quantum walk can be performed on the OAM space of a laser beam and, therefore, we don't need single photons. 

Homogenously polarized optical fields can be treated as scalar fields under the paraxial approximation. In the general case the optical field can be represented as the sum of two orthogonally polarized fields
\begin{align}
\bm{E}(\bm{r},z) &=   E_R(\bm{r},z) \hat{\bm{e}}_R + E_L(\bm{r},z) \hat{\bm{e}}_L =E_0 {\left[\begin{array}{c} u_{_R}(\bm{r},z) \\ u_{_L}(\bm{r},z) \end{array}\right]}\label{eqn-14},
\end{align}
where $E_R(\bm{r},z)$ and $E_L(\bm{r},z)$ are the electric field components for right-handed and left-handed circular polarization, respectively, $\bm{r}$ is the two-dimensional position vector on the transverse plane perpendicular to the propagation direction ($z$-axis), and $\hat{\bm{e}}_R$ and $\hat{\bm{e}}_L$ are the unit vectors associated with right-handed and left-handed circular polarization. In the last line of Eq.~\eqref{eqn-14} we express the optical field as a Jones vector, times an overall amplitude $E_0$. %For convenience we'll express the Jones vectors (and Jones matrices) in the circular polarization basis.}
%The norm of the Jones vector in the last line of Eq.~\eqref{eqn-14} is the square root of the intensity 
%\FSR{The expression $|E_0|^2 = |E_R|^2 + |E_L|^2$ is not correct: $E_0$ is a constant, the other two are functions.} 

The two position dependent components of the Jones vector can now be expanded in terms of any complete set of orthogonal modes. We use the Laguerre-Gaussian modes, which are OAM eigenstates and solutions of the paraxial wave equation in cylindrical coordinates \cite{Allen1992,Leach2002}. These modes are distinguished by two integer indices, an azimuthal index $\ell$, which is proportional to the OAM of the mode, and a positive radial index $p$. Not being much concerned about the radial dependence and to simplify the analysis, we evaluate the sum over $p$. The components of the Jones vector are then given by
\begin{align} 
u_{L,R}(\bm{r},z)& = u_{L,R}(r,\phi,z) \nonumber\\ &= \sum_{\ell} a_{\ell}^{(L,R)}(r) \exp(i\ell\phi) \exp(-i k z) , 
\end{align} 
%\begin{align} u_{L,R}(\bm{r},z)& = u_{L,R}(r,\phi,z) \nonumber\\ &= \sum_{p,\ell} \alpha_{p,\ell}^{(L,R)} f_{p,\ell}(r,z) \exp(i\ell\phi) \exp(-i k z) , \end{align}
where $a_{\ell}^{(L,R)}(r)$ represents radial-dependent expansion coefficients 
%$f_{p,\ell}(r,z)$ is the radial function of the mode, $\ell$ is the azimuthal index (an integer), which is proportional to the OAM of the mode, $p$ is a radial index (positive integer) 
and $k$ is the wave number. For convenience we  set $z=0$ from now on and consider the optical field in the image plane, purely as a function of $r$ and $\phi$. 
%Not being much concerned about the radial dependence and to simplify the analysis, we evaluate the sum over $p$ to obtain
%\begin{align} u_{L,R}(r,\phi) &= \sum_{\ell} a_{\ell}^{(L,R)}(r) \exp(i\ell\phi) , \end{align} 
%where $a_{\ell}^{(L,R)}(r) = \sum_{p} \alpha_{p,\ell}^{(L,R)} f_{p,\ell}(r,0)$. 
The normalization of the Jones vector implies that 
\begin{align}
2\pi \int \sum_{\ell} \left[ \left| a_{\ell}^{(L)}(r) \right|^2 +  \left| a_{\ell}^{(R)}(r) \right|^2 \right]\ r{\rm d}r = 1 .
\end{align}  

The action of an optical device on the state of light is represented by a $2\times 2$  Jones matrix. For a quarter-wave plate in the circular polarization basis it is given by \cite{Hecht2001} 
\begin{align}
J_w &= \frac{1}{\sqrt{2}}\left( \begin{array}{cc} 1 & i \\ i & 1 \end{array}\right) .
\end{align} 

A $q$ plate causes spin-orbital coupling --- turning spin angular momentum into OAM.
%It acts like a half-wave plate, with varying orientation of its optic axis over the transverse plane. 
% \TK{Omitted a sentence.}
At each point on the transverse plane the $q$ plate behaves like a half-wave plate with a particular orientation of its optics axis. The orientation of the optic axis is given in terms of the azimuthal angle $\phi$ --- its orientation angle is $q\phi$, where $q$ is a half integer
%collection of many small half-wave plates that are rotated such that the orientation of their optic axes vary over the transverse plane of the $q$ plate. A half-wave plate turns left-handed circular polarization into right-handed circular polarization and vice versa. Let's assume that the optic axis of each half-wave plate make\FSRnew{s} an angle \FSRnew{of} $q\phi$ with the direction of horizontal polarization, where $q$ is a half-integer. %and $\phi$ is the azimuthal angle of the wave plate's position. 
%\FSR{The previous explanation was a little confusing. It seemed as if there are several distinct half-wave plates involved.} 
Then the combined action of the rotated half-wave plates can be expressed concisely by a $\phi$-dependent Jones matrix
\begin{align}
J_q &= R(-q\phi)\left(\begin{array}{rr}
0 & 1  \\ 
1 & 0 \end{array}\right)R(q\phi) = \left(\begin{array}{cc}
0 & e^{-i2q\phi}  \\ 
e^{i2q\phi} & 0 \end{array}\right),
\end{align}
where $R(q\phi)$ is the rotation matrix ${\rm diag}(e^{iq\phi}, e^{-iq\phi})$. %that is diagonal with respect to the circular polarization basis.
%\begin{align}
%R(q\phi) &= \left(\begin{array}{cc}
%e^{iq\phi} & 0  \\ 
%0 & e^{-iq\phi} \end{array}\right).
%\end{align}

The concatenation of a $q$ plate and a 45 degree rotated quarter-wave plate, leads to
\begin{align}
J_w\left(\frac{\pi}{4}\right) J_q &= \frac{1}{\sqrt{2}} \left(\begin{array}{cc} 1 & 1 \\ -1 & 1\end{array}\right) 
\left(\begin{array}{cc} 0 & e^{-i2q\phi}  \\ e^{i2q\phi} & 0 \end{array}\right) \nonumber\\
& = \frac{1}{\sqrt{2}} \left(\begin{array}{cc} e^{i2q\phi} & e^{-i2q\phi} \\ e^{i2q\phi} & -e^{-i2q\phi} \end{array}\right) \nonumber\\ 
& = \frac{1}{\sqrt{2}} \left(\begin{array}{cc} 1 & 1 \\ 1 & -1 \end{array}\right) 
\left(\begin{array}{cc} e^{i2q\phi} & 0\\ 0 & e^{-i2q\phi}\end{array}\right) \label{eq-18}.
\end{align}
We see that the action of the diagonal matrix $\mbox{diag}(e^{i2q\phi},e^{-i2q\phi})$ on the OAM eigenstates is the following: the beam with right-handed circular polarization will gain OAM of $2q\hbar$ per photon and the one with left-handed circular momentum will lose OAM of $2q\hbar$ per photon. Hence, it  represents the conditional shift operator in a quantum walk. The first  matrix in Eq.~\eqref{eq-18} corresponds to the Hadamard operator Eq.~(\ref{Hadamard}). Hence, the $q$ plate and the rotated quarter-wave plate together produce the same quantum walk evolution as $\tilde{H}S$. If the initial state has an azimuthal index of $\ell=0$ and its state of polarization is $(\hat{\bm{e}}_R+i\hat{\bm{e}}_L)/\sqrt{2}$, the state after the first iteration reads 
\begin{align}
\bm{E}(\bm{r}) = \frac{E_0\alpha_{0}(r)}{2} &\left[e^{i\phi} \left(\hat{\bm{e}}_R+\hat{\bm{e}}_L\right)+ ie^{-i\phi} \left(\hat{\bm{e}}_R-\hat{\bm{e}}_L\right)\right] ,
\label{entanglement2}
\end{align}
in analogy to the state (\ref{entanglement1}).

%\section{Implementation scheme}
%\label{imple}

\begin{figure}
\includegraphics[width=7.5cm]{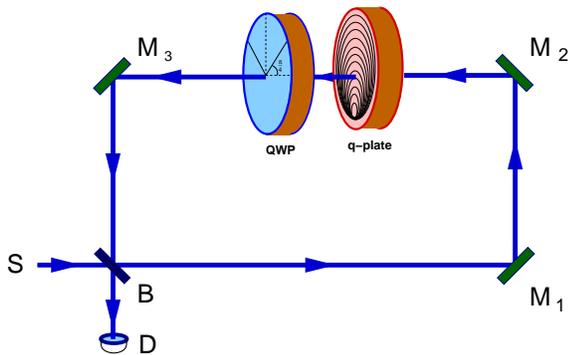}
\caption{Diagrammatic representation of the optical setup of a quantum walk. A laser beam passes through a $q$ plate, followed by a quarter-wave plate. The loop corresponds to a single iteration of a quantum walk process.}
\label{scheme-01}
\end{figure}

The setup to realize the quantum walk with laser light is depicted in Fig.~\ref{scheme-01}. A laser pulse with zero OAM and a circular state of polarization is sent through beam splitter $B$, with a transmission coefficient $\mu$. Upon entering the ring interferometer the beam is reflected from mirrors $M_1$ and $M_2$ and passes through the $q$ plate, %\footnote{A $q$ plate with q=1/2 can, for instance be manufactured using form birefringence in sub-wavelength gratings (see for instance %\cite{niv}). In experimental tests of the OAM modes  created by such a $q$ plate from a Gaussian beam ($\ell=0$) we found no measurable distortion}
followed by the 45 degree rotated quarter-wave plate. After reflection from mirror $M_3$, a fraction $\mu$ of the beam is transmitted through beam spitter $B$ to the detector $D$. The remaining fraction of the beam stays in the ring interferometer. \TK{In order to avoid beam divergence and phase mismatch (for example due to the Gouy phase shift \cite{saleh2007}) a 4-f imaging system can be incorporated into the loop which maps the plane of the $q$ plate to itself. } Each round trip represents one iteration of the DTQW process. 
Based on experimental tests of $q$ plates with $q=1/2$ \cite{niv2} a hundred iterations 
%seem to
would be feasible 
%for a sufficiently
with a reasonably intense light source.
%Based on our preliminary experimental tests of the mode fidelity of a $q$ plate with $q=1/2$ and the fact that losses in the ring cavity can be compensated by choosing an appropriate initial laser intensity a quantum walk of a hundred steps is not unrealistic.}  
The measurement of the light coupled out in each iteration yields the intensity distribution over the OAM basis states. 

The detection of the OAM spectrum can be done with the aid of an efficient OAM sorter \cite{lavery2012}, which employs refractive optical elements to perform a conformal mapping of the optical field, turning the azimuthal phase variation into a linear phase variation. A subsequent Fourier transform that is implemented by a lens will separate different OAM orders as adjacent bright spots, which can be measured simultaneously on a CCD array to reveal an OAM spectrum of up to  a hundred $\ell$ values \cite{Dudley2013}. This bandwidth can be doubled by separating  the output into odd and even OAM values first \cite{Lavery.et.al.2011}. A further enhancement of the bandwidth is possible by  making separate measurements on different parts of the spectrum, centered, for example, around $\ell=0$, $\ell=100$ and $\ell=-100$, respectively.
%measuring different parts of the OAM spectrum {\sl sequentially}}.   

%\section{Discussion}
%\label{coherent}

Above, we proposed two different realizations of a coined quantum walk in the OAM spectrum for (i) a single photon and (ii) a laser pulse and discussed their physical implementation. In both cases, we found that, after the first iteration, the optical field cannot be factored into a product of the spatial beam profile and the state of polarization. For the single photon this structure [Eq.~(\ref{entanglement1})] implies quantum entanglement. The nonfactorizability of the electromagnetic field Eq.~(\ref{entanglement2}) has been called `nonquantum entanglement' \cite{Simon2010}. The formal analogy between both forms of entanglement was used \cite{Simon2010} to show that in classical optics those Mueller matrices that represent positive but not completely positive operations are not physically realizable. Here we identified another application of nonquantum entanglement, namely a coined quantum walk with OAM modes of light.

%In the remainder of this article 
Next we demonstrate, using the quantum description of laser light, that nonfactorizable classical optical fields Eq.~(\ref{entanglement2}) do not contain quantum entanglement. To this end we represent the state of the laser beam as a pure coherent state $\ket{\alpha}$ with two additional indices $s$ for polarization and $\ell$ for OAM:
\begin{align}
\ket{\alpha, s, \ell} &= \exp(\alpha a^\dagger_{s,\ell} - \alpha^* a_{s,\ell})\ket{0},
\end{align}
where $a^\dagger_{s,\ell}$ is the mode creation operator, which creates a photon in the $s$ polarization state with an OAM of $\ell \hbar$.

The action of the $q$ plate and the quarter-wave plate on the mode creation operators is given by
\begin{align}
\mbox{$q$ plate:}\qquad a^\dagger_{s,\ell} &\rightarrow a^\dagger_{\bar{s},\ell\pm 1},\\
\mbox{QWP: }\qquad a^\dagger_{R,\ell} &\rightarrow \frac{1}{\sqrt{2}}\left(a^\dagger_{R,\ell} - a^\dagger_{L,\ell}\right),\\
a^\dagger_{L,\ell} &\rightarrow \frac{1}{\sqrt{2}}\left(a^\dagger_{R,\ell} + a^\dagger_{L,\ell}\right) ,
\end{align}
where $\bar{s}$ represents the state of polarization opposite to $s$. Hence, the combined action of the $q$ plate, followed by the quarter-wave plate on the coherent state reads
%\begin{align}
%a^\dagger_{R,\ell} &\rightarrow \frac{1}{\sqrt{2}}\left(a^\dagger_{R,\ell+ 1} + a^\dagger_{L,\ell+ 1}\right) \\
%a^\dagger_{L,\ell} &\rightarrow \frac{1}{\sqrt{2}}\left(a^\dagger_{R,\ell- 1} - a^\dagger_{L,\ell- 1}\right) .
%\end{align}
%This reflects in the coherent states as
\begin{align}
\ket{\alpha, R,\ell} &\to \ket{\frac{\alpha}{\sqrt{2}}, R,\ell+1}\otimes\ket{\frac{\alpha}{\sqrt{2}}, L,\ell+1},\\
\ket{\alpha, L,\ell} &\to \ket{\frac{\alpha}{\sqrt{2}}, R,\ell-1}\otimes\ket{-\frac{\alpha}{\sqrt{2}}, L,\ell-1} ,
\end{align}
%From here we can see 
which represents the quantum walk evolution in terms of coherent states. Consider for example, an initial coherent state with an OAM of $\ell=0$ and with a polarization state given by $(\ket{R} + i\ket{L})/\sqrt{2}$. This state is represented by the product
\begin{align}
\ket{\psi_0} = \ket{\frac{\alpha}{\sqrt{2}}, R, 0}\otimes \ket{\frac{i\alpha}{\sqrt{2}}, L, 0} .
\end{align}
Then after the first quantum walk iteration the state becomes
\begin{align}
\ket{\psi}_0 \to \ket{\psi}_1 = & \ket{\frac{\alpha}{2}, R, 1} \otimes\ket{\frac{\alpha}{2}, L, 1} \nonumber \\ & \otimes \ket{i\frac{\alpha}{2}, R, -1} \otimes\ket{-i\frac{\alpha}{2}, L, -1} ,
\end{align}
which is clearly a product state without quantum entanglement.

In summary, we propose a new method to implement a quantum walk, using laser light in an optical setup that contains a quarter-wave plate and a $q$ plate in a ring interferometer. The advantage of this implementation is that it  neither requires potentially unstable beam splitters arrays, nor sources and detectors for single photons. Moreover, the results of an arbitrary number of iterations are obtained as a real-time sequence at the output. The fact that we use a laser source implies that the output state is not quantum entangled, but contains so-called nonquantum entanglement which enables the coined quantum walk. The extent to which bright, OAM carrying, classical light can be used to simulate quantum computing or dynamics of quantum systems by means of quantum walks \cite{Lovett.et.al2010, Schreiber.et.al2012} is a question for future research. Note, that our scheme allows to observe the Hong-Ou-Mandel effect \cite{Hong1987, Nagali2009} without a spatial interferometer \footnote{A photon entering in the state with OAM value $\ell$ and right circular polarization together with a photon with OAM value $\ell+2$ and left-circular polarization 
%$\ket{l,R}\ket{l+2,L}$ 
are transformed after two roundtrips in the setup into { a state of two photons with OAM value $\ell$ in a superposition with two photons with OAM value $\ell+2$}. This photon bunching resembles the Hong-Ou-Mandel effect. { It is experimentally observed \cite{Nagali2009}
} and shows that it is not possible to simulate all aspects of the dynamics of many-particle quantum walks with classical light.}. 

T.K. acknowledges the partial support from the National Research Foundation of South Africa [Grant No. 86325 (UID)].
%\bibliography{ref_03}
%\begin{thebibliography}
%\input{qwalknew.bbl}
%\end{thebibliography}
%Merlin.mbs v4.21 2009-07-09.
%

\end{document}